%% file: SpinOneSOC.tex
\documentclass{nature}

\usepackage{color, units, graphicx, amsmath, amssymb, bm, float}
\include{macros}

\title{Itinerant magnetism in spin-orbit coupled Bose gases}

\author{D. L. Campbell, R. M. Price, A. Putra, A. Vald\'{e}s-Curiel, D. Trypogeorgos, and I. B. Spielman}

\begin{document}
\maketitle

\begin{affiliations}
 \item Joint Quantum Institute, University of Maryland, College Park, Maryland,
20742, USA
\end{affiliations}

\input{Introduction}

\input{PhaseTransition}

\input{Metastable}


\input{Conclusion}

\input{MethodsSummary}

\bibliography{SpinOneSOC}


\begin{addendum}
\item This work was partially supported by the ARO's atomtronics MURI, by the AFOSR's Quantum Matter MURI, NIST, and the NSF through the PFC at the JQI.
 
\item[Author Contributions] All authors excepting I.B.S contributed to the data taking effort.  All authors: analyzed data; performed numerical and analytical calculations; and contributed to writing the manuscript.  I.B.S. proposed the initial experiment (with great enthusiasm, but with little appreciation for the many technical hurdles to be resolved by the remainder of the team).

\item[Competing Interests]  The authors declare that they have no competing financial interests.

\item[Correspondence] Correspondence and requests for materials should be addressed to I.B.S.~ (email: ian.spielman@nist.gov).

\end{addendum}

\end{document}

%% file: macros.tex
\newcommand{\onlinecite}[1]{\nocite{#1}\citenum{#1}}

\def\ms{{\ {\rm ms}}}						

\def\Er{{{E_R}}}							
\def\kr{{{k_R}}}							
\def\Rb87{^{87}\rm{Rb}}					

\def\ex{{\mathbf e}_x}                            
\def\ey{{\mathbf e}_y}                            
\def\shorteq{\! = \!}                            
\DeclareMathAlphabet\mathbfcal{OMS}{cmsy}{b}{n}

\newcommand{\ket}[1]{|#1\rangle}


%% file: Introduction.tex


\begin{abstract}
Phases of matter are conventionally characterized by order parameters describing the type and degree of order in a system.  For example, crystals consist of spatially ordered arrays of atoms, an order that is lost as the crystal melts.  Likewise in ferromagnets, the magnetic moments of the constituent particles align only below the Curie temperature, ${\bm T_{\rm C}}$.  These two examples reflect two classes of phase transitions: the melting of a crystal is a first-order phase transition (the crystalline order vanishes abruptly) and the onset of magnetism is a second-order phase transition (the magnetization increases continuously from zero as the temperature falls below ${\bm T_{\rm C}}$).  Such magnetism is robust in systems with localized magnetic particles, and yet rare in model itinerant systems\cite{Stoner1933,Sanner2012} where the particles are free to move about.  Here for the first time, we explore the itinerant magnetic phases present in a spin-1 spin-orbit coupled atomic Bose gas\cite{Lin2011,Lan2014,Radic2014,Hickey2014}; in this system, itinerant ferromagnetic order is stabilized by the spin-orbit coupling, vanishing in its absence.  We first located a second-order phase transition that continuously stiffens until, at a tricritical point, it transforms into a first-order transition (with observed width as small as ${\bm h\times4\ {\rm Hz}}$).  We then studied the long-lived metastable states associated with the first-order transition.  These measurements are all in agreement with theory. 
\end{abstract}

Most magnetic systems are composed of localized particles such as electrons\cite{Aharoni2011}, atomic nuclei\cite{Abragam1992}, and ultracold atoms in optical lattices\cite{Kronjager2010,Simon2011,Hart2014}, each with a magnetic moment ${\boldsymbol \mu}$.  By contrast, itinerant magnetism\cite{Stoner1933} describes systems where the magnetic particles -- here ultracold neutral atoms -- can themselves move freely, and for which magnetism is generally weak.  Our spin-orbit coupled Bose-Einstein condensates (BECs) constitute a magnetically ordered itinerant system in which the atoms' kinetic energy explicitly drives a phase transition between two different ordered phases\cite{Lan2014}.  The coupling between spin and momentum afforded by spin-orbit coupling (SOC) provides a new route for stabilizing ferromagnetism in itinerant systems, both for fermions\cite{Zhang2014} and here for bosons\cite{Radic2014,Hickey2014}.

\begin{figure*}
\begin{center}
\includegraphics{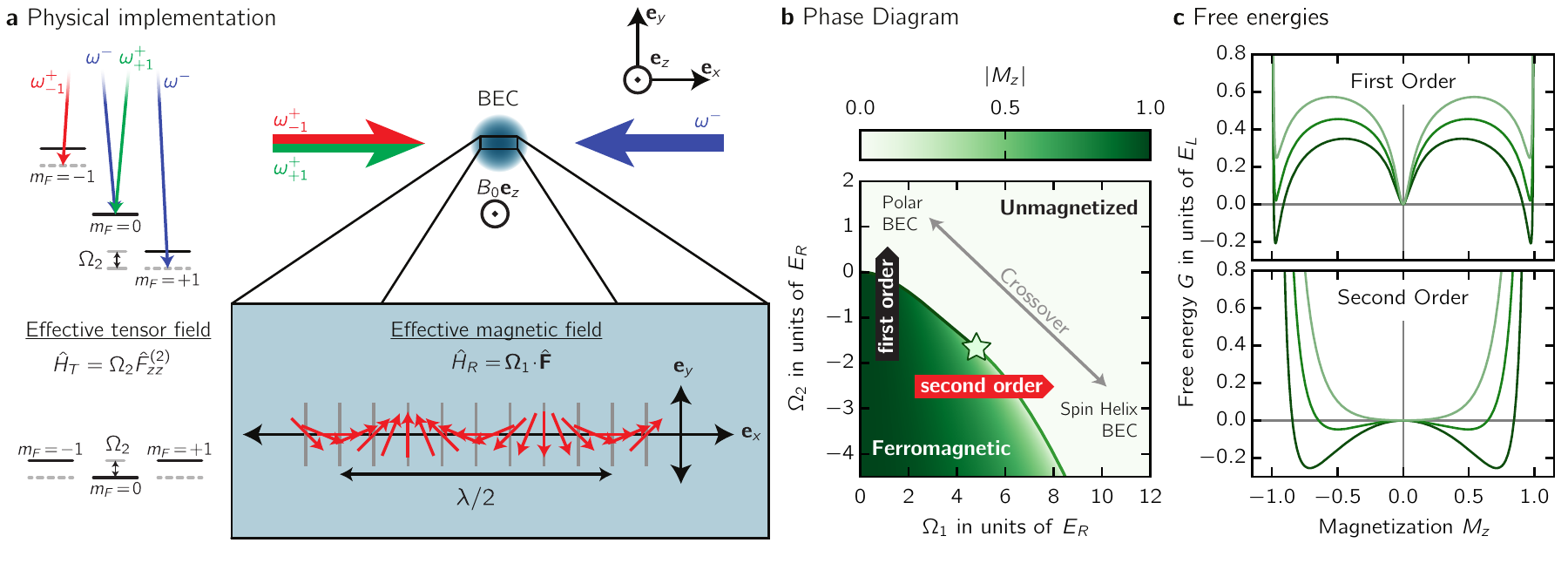}
\caption{{\bf Experimental system}
{\bf a} Schematic and level diagram.  The $\ket{m_F = -1}\leftrightarrow\ket{m_F = 0}$ and $\ket{m_F = 0}\leftrightarrow\ket{m_F = +1}$ transitions of the $f=1$ ground state manifold of $^{87}{\rm Rb}$ were independently Raman coupled, giving experimental control of $\Omega_1$ and $\Omega_2$.  {\bf b} Phase diagram.  The ferromagnetic order parameter $\left|M_z\right|$ is plotted against $\Omega_2$ and $\Omega_1$.  The solid (dashed) red curve denotes the first-order (second-order) transition from the magnetized phase.  {\bf c} Free energies.  Top:  near the first-order phase transition at $\Omega_1/\Er = 1$ for $\Omega_2/\Er = -0.35, -0.1$ and $0.15$ for the black, blue and red traces respectively, as marked by the red flags in {\bf b}.  Bottom: near the second-order phase transition at $\Omega_2/\Er = -2.5$ for $\Omega_1/\Er = 4.5, 5.5$, and $6.5$ for the black, blue and red traces respectively.}
\label{Fig:Setup}
\end{center}
\end{figure*}

Phase transitions can generally be described in terms of a free energy $G(M_z)$ -- including the total internal energy along with thermodynamic contributions -- that is minimized for an equilibrium system.  Here the magnetization $M_z=\langle \hat S_z\rangle/\hbar$ is an order parameter which changes abruptly as our system undergoes a phase transition, where $\hat {\bf S}$ is the spin.  Figure~\ref{Fig:Setup}{\bf c} shows typical free energies: a first-order phase transition (top panel) occurs when the number of local minima in $G(M_z)$ stays fixed, but the identity of the global minima changes; and a second-order phase transition (bottom panel) occurs when degenerate global minima merge or separate.  These defining features are true both for $T>0$ thermal and $T=0$ quantum phase transitions.  Here we realize $f=1$ BECs with SOC and study the magnetic order and associated quantum phase transitions present in their phase diagram.

For spin-1/2 systems (i.e, total angular momentum, $f=1/2$) like electrons, ferromagnetic order can be represented in terms of a magnetization vector ${\bf M} = \langle\hat {\bf S}\rangle/\hbar$.  This is rooted in the fact that the three components of the spin operator $\hat {\bf S}$ transform vectorially under rotation.  More specifically, any Hamiltonian describing a two level system may be expressed as $H = \hbar\Omega_0 + {\bf \Omega}_1\!\cdot\!\hat{\bf S}$, the sum of a scalar (rank-0 tensor) and a vector (rank-1 tensor) contribution.  The former, described by $\Omega_0$ gives an overall energy shift, and the latter takes the form of the linear Zeeman effect from an effective magnetic field proportional to ${\bf \Omega}_1$.  Going beyond this, fully representing a spin-1 (i.e., total angular momentum $f=1$) Hamiltonian with angular momentum $\hat {\bf F}$ rather than spin $\hat {\bf S}$, requires an additional five-component rank-2 tensor operator -- the quadrupole tensor -- and therefore there exist ``magnetization'' order parameters that are not simply associated with any spatial direction\cite{Stenger1998,Barnett2006,Stamper-Kurn2013}.

Pioneering studies in GaAs quantum wells\cite{Stanescu2007a,Koralek2009} showed that material systems with equal contributions of Rashba and Dresselhaus SOC  described by the term $2\hbar\kr k_x \hat F_z / m$, subject to a transverse magnetic field with Zeeman term $\Omega_1 \hat F_x$, can equivalently be described as a spatially periodic effective magnetic field.  
Our spin-orbit coupled spin-1 atomic systems\cite{Wang2010,Lin2011,Ho2011,Zhang2012,Wang2012a,Cheuk2012,Parker2013,Galitski2013} have this form of SOC and can therefore be described by the magnetic Hamiltonian
\begin{align}
\hat H &= \frac{\hbar^2{\bf k}^2}{2m} + {\boldsymbol\Omega}_1(x)\!\cdot\!\hat{\bf F} + \Omega_2\hat F^{(2)}_{zz}, \label{eq:Magnetic}
\end{align}
describing atoms with mass $m$ and momentum $\hbar{\bf k}$ interacting with an effective Zeeman magnetic field ${\boldsymbol\Omega}_1(x)/\Omega_1 = \cos(2 \kr x)\ex - \sin(2 \kr x)\ey$ helically precessing in the $\ex$-$\ey$ plane with spatial period $\pi/\kr$ set by the SOC strength; and an additional Zeeman-like tensor coupling with strength $\Omega_2$.   Here, $\hat F^{(2)}_{zz}/\hbar = \hat F_z^2/\hbar^2-2/3$ is an element of the quadrupole tensor operator.  

The competing contributions -- from kinetic and magnetic ordering energies -- make ours an archetype system for studying robust itinerant magnetic order and understanding the associated phase transitions, of which both first- and second- order are present (Fig.~\ref{Fig:Setup}{\bf b}).  For infinitesimal $\Omega_1$, the tensor field favors either: for $\Omega_2>0$, a polar BEC ($m_F=0$: unmagnetized, $M_z = 0$), or for $\Omega_2<0$, a ferromagnetic BEC ($m_F = +1$ or $-1$: magnetized, $|M_z| = 1$); these phases are separated by a first-order phase transition at $\Omega_2 = 0$.  In contrast, for large $\Omega_1$ the system forms a spin helix BEC (with local magnetization antiparallel to ${\boldsymbol\Omega}_1$: unmagnetized, $M_z = 0$).  This order increases the system's kinetic energy, leading to the second-order phase transition shown in Fig.~\ref{Fig:Setup}{\bf b}.  These two extreme limits continuously connect at the point $(\Omega_1^*,\Omega_2^*)$, the green star in Fig.~\ref{Fig:Setup}{\bf b}, where the small-$\Omega_1$ first-order phase transition gives way to the large-$\Omega_1$ second-order transition, and together these regions constitute a curve of critical points $\{(\Omega_1^{C},\Omega_2^{C})\}$.

As shown in Fig.~\ref{Fig:Setup}{\bf a}, we realized this situation by illuminating $\Rb87$ BECs in the total angular momentum $f=1$ ground state manifold with a pair of counter propagating ``Raman'' lasers that coherently coupled the manifold's three $m_F$ states.  As we first showed\cite{Lin2011} using effective $f=1/2$ systems, this introduces both a spin-orbit and a Zeeman term into the BEC's Hamiltonian, equivalent to Eq.~\eqref{eq:Magnetic}.  The Raman laser wavelength $\lambda$ defines the natural units for our system: the single-photon recoil energy $\Er = \hbar^2 \kr^2 / 2 m$, and momentum $\hbar \kr = 2\pi\hbar/\lambda$ setting the SOC strength.  Here the quadratic Zeeman shift from a large bias magnetic field $B_0{\bf e}_z$ split the low-field degeneracy of the $\ket{m_F = -1}\leftrightarrow\ket{m_F = 0}$ and $\ket{m_F = 0}\leftrightarrow\ket{m_F = +1}$ transitions, and we independently Raman coupled these state-pairs with equal strength $\Omega_1$.  As described in the Methods Summary, we dynamically tuned the quadrupole tensor field strength $\Omega_2$ by simultaneously adjusting the Raman frequency differences.  Without this technique, only the upper half-plane of the phase diagram (Fig.~\ref{Fig:Setup}{\bf b}) would be accessible: containing only an unmagnetized phase, therefore lacking any phase transitions. 

In each experiment, we first prepared BECs at a desired point in the phase diagram, possibly having crossed the phase transition during preparation; a combination of trap dynamics\cite{Lin2009a,Lin2011a}, collisions, and evaporation\cite{Ji2014} kept the system in or near (local) thermal equilibrium.  We then made magnetization measurements directly from the spin resolved momentum distribution obtained using the time-of-flight techniques described in Ref.~\onlinecite{Lin2011a}.

%% file: PhaseTransition.tex

\begin{figure*}
\includegraphics{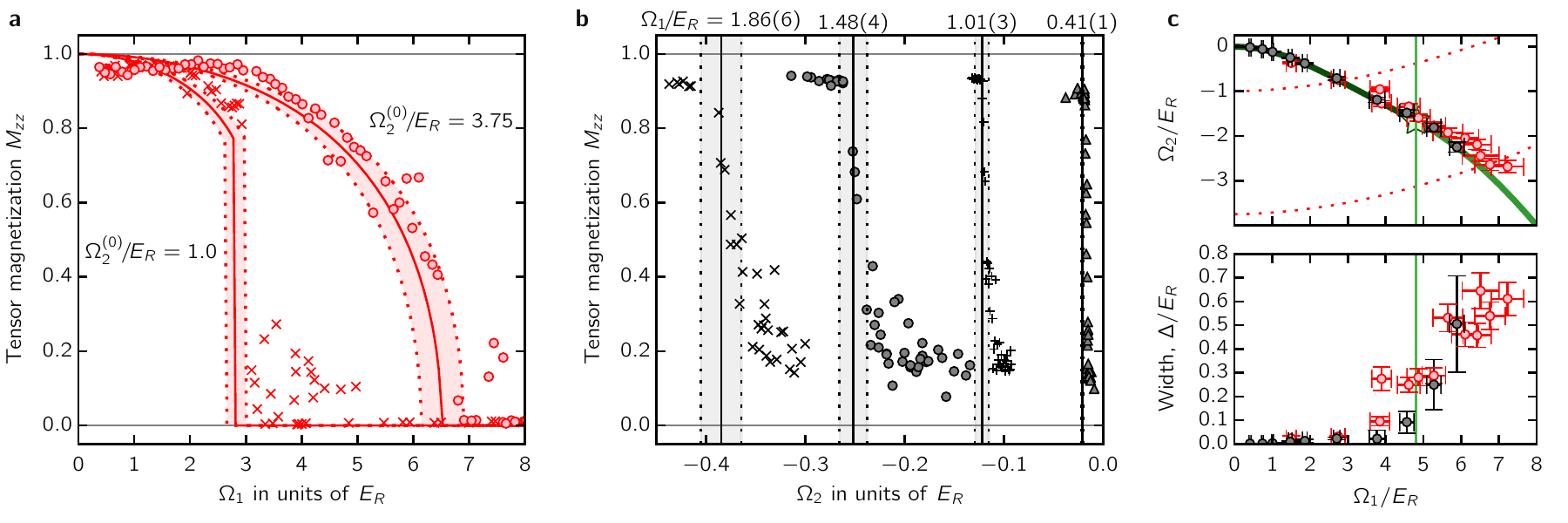}
\caption{{\bf Measured phase transition}  {\bf a}  Tensor magnetization $M_{zz}$ measured as a function of $\Omega_1$, showing both second-order [$\Omega_2(\Omega_1=0) = -2.0000(3) E_{R}$]  and first-order [$\Omega_2(\Omega_1=0) = -1.0 E_{R}$] phase transitions in comparison with theory.  These curves followed the nominally horizontal trajectories (see Methods Summary) marked by red dashed curves in {\bf c}. {\bf b} Tensor magnetization measured as a function of $\Omega_2$ at $\Omega_1/E_R = 1.86(6)$, $1.48(4)$, $1.01(3)$, and $0.41(1)$, plotted along with the predicted critical $\Omega_2$.  In {\bf a} and {\bf b} the light-colored region reflects the uncertainty in theory resulting from our $\approx5\%$ systematic uncertainty in $\Omega_1$. {\bf c} Phase transition.  Black (red) symbols depict data obtained using vertical (nominally horizontal) cuts through the phase diagram.  Top, measured phase transitions plotted along with theory: solid (phase transition), and green vertical line (tricritical point, $\Omega_1^*$).   Bottom: 20\% to 50\% transition width showing the clear shift from first to second-order with increasing $\Omega_1$. 
}\label{Fig:SecondOrder}
\end{figure*}

Our experiment first focused on thermodynamic phase transitions.  We made vertical (horizontal) scans through the phase diagram by initializing the system in the unmagnetized phase at a desired value of $\Omega_1$ ($\Omega_2$) with $\Omega_2\gtrsim0$ ($\Omega_1 \lesssim 10 \Er$), and then ramping $\Omega_2$ ($\Omega_1$) through the transition region.  (As discussed in the methods summary our nominally horizontal $\Omega_1$-scans followed slightly curved trajectories through the phase diagram, such as the red dashed curve in Fig.~\ref{Fig:SecondOrder}c).  Following such ramps, domains with both $\pm M_z$ can rapidly form, and we therefore focus on the tensor magnetization $M_{zz} = \langle \hat F_{zz}^{(2)} \rangle / \hbar + 2/3$ which is sensitive to this local magnetic order.

Using horizontal scans, we crossed through the second-order phase transition ($\Omega_2 < \Omega_2^*$) where the free energy evolves continuously from having one minimum (with $M_{zz}=0$, for large $\Omega_1$) to having two degenerate minima (with $M_{zz}>0$, for smaller $\Omega_1$).  As shown in Fig.~\ref{Fig:SecondOrder}{\bf a}, $M_{zz}$ continuously decreases, reaching its saturation value as $\Omega_1$ exceeds $\Omega_1^C$.  Repeating this processes for $\Omega_2^*<\Omega_2<0$, we found a sharp first-order transition.  In each case, data is plotted along with theory with no adjustable parameters.  Using data of this type for a range of $\Omega_2$ and fitting to numeric solutions of Eq.~(\ref{eq:Magnetic}), we obtained the critical points plotted in Fig. \ref{Fig:SecondOrder}{\bf c}.  Because horizontal cuts through the phase diagram are nearly tangent to the transition curve for small $\Omega_2$, this produced large uncertainties in $\Omega_1^C$ for the first-order phase transition.

We studied the first-order phase transition with greater precision by ramping $\Omega_2$ through the transition at fixed $\Omega_1$ (Fig.~\ref{Fig:SecondOrder}{\bf b}) and found near perfect agreement with theory.  For all the experimentally measured critical points, see Fig. \ref{Fig:SecondOrder}c top, separating the unmagnetized and ferromagnetic phase, we also measured the corresponding transition widths.  Defined as the required interval for the curve to fall from 50\% to 20\% of its full range, this width $\Delta$ decreases sharply at $\Omega_1^*$, marking the crossover between second- and first-order phase transitions (see Fig.~\ref{Fig:SecondOrder}c bottom).  In these data, the width of the first-order transition becomes astonishingly narrow: as small as $0.0011(3)\Er = h\times4(1) {\rm Hz}$ at $\Omega_1 = 0.41(1)$.

%% file: Metastable.tex

\begin{figure}
\includegraphics{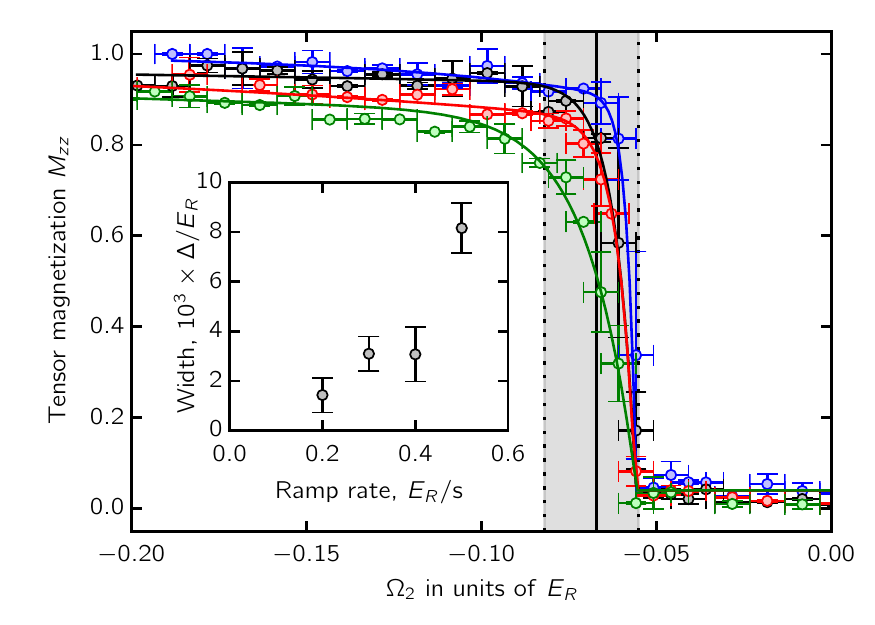}
\caption{{\bf Quenching dynamics} The system was prepared in the unmagnetized phase with $\Omega_1 = 0.74(8) E_R$ and $\Omega_2$ was ramped through the phase transition at ramp-rates $d\Omega_2/dt = -0.2, -0.3, -0.4$, and $-0.5 E_R/{\rm s}$ (blue, black, red, and green symbols, respectively).  The inset shows the decreasing width of the first-order transition as the ramp-rate decreases.}
\label{Fig:quench}
\end{figure}

We observed that scans crossing the second-order transition typically required $50\ms$ to equilibrate, while for scans crossing the first-order transitions we allowed as long as $1.5\ {\rm s}$ for equilibration.  Systems taken through a first-order phase transition can remain in long-lived metastable states.  Here a metastable state with $M_z = 0$ persists in the ferromagnetic phase, and a pair of metastable states with $M_z \neq 0$ persists in the unmagnetized phase.  We began our study of this metastability by quenching through the first-order transition at $\Omega_1 = 0.74(8)\Er$ with differing rates from $0.5$ to $0.2\ \Er/{\rm s}$, as shown in Fig.~\ref{Fig:quench}.  We observed the transition width continuously decreases with decreasing ramp rate (inset to Fig.~\ref{Fig:quench}), consistent with slow relaxation from a metastable initial state. 

\begin{figure}
\includegraphics{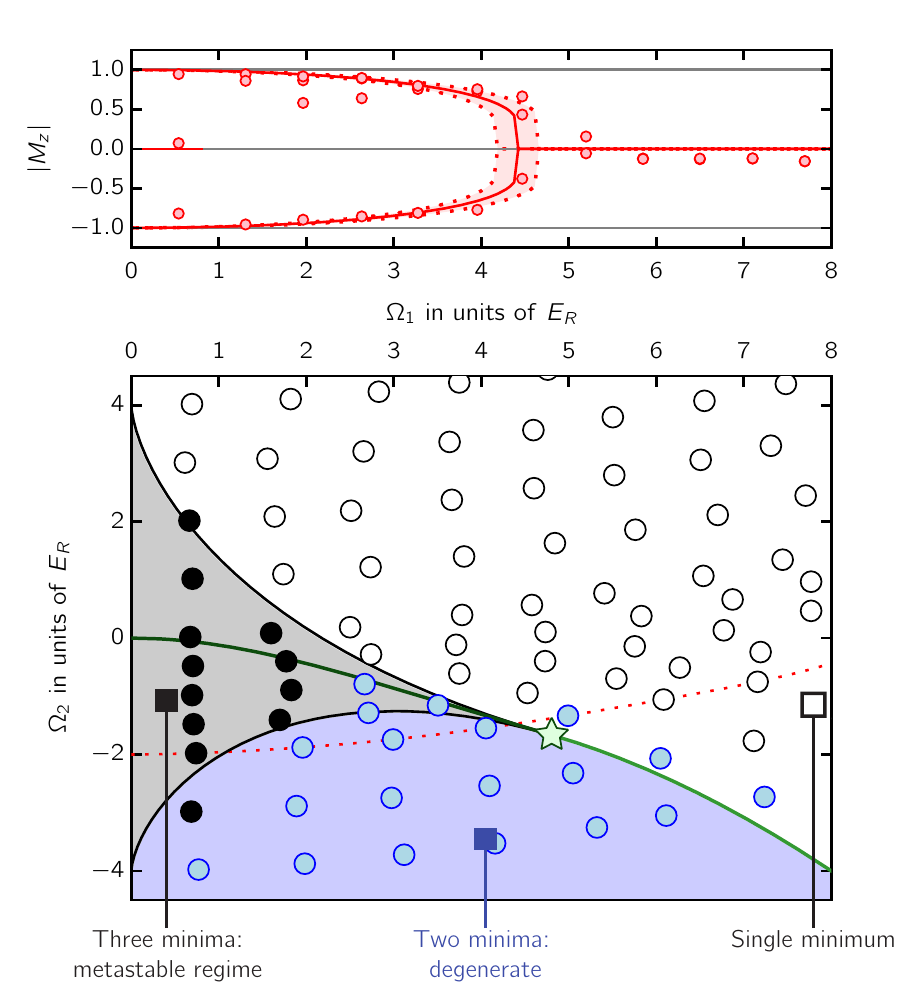}
\caption{{\bf Metastable states}  Top, Measured magnetization plotted along with theory.  The system was prepared at the desired $\Omega_2 = -2 E_{R}$; $\Omega_1(t)$ was then increased to its displayed final value; during this ramp $\Omega_2$ also changed, and the system followed the curved trajectory in the bottom panel. Each displayed data point is an average of up to 15 measurements, and the colored region reflects the uncertainty in theory resulting from our $\approx5\%$ systematic uncertainty in $\Omega_1$. Bottom, state diagram: theory and experiment. Blue: two states; black: three states; white: one state.  Colored areas denote calculated regions where the color-coded number of stable/metastable states are expected.  Symbols are the outcome of experiment. Each displayed data point is an average of up to 20 measurements.}
\label{Fig:metastable}
\end{figure}

We explored the full regime of metastability by initializing BECs in each of the $\ket{m_F = 0,\pm1}$ states, at fixed $\Omega_2$, then rapidly ramping $\Omega_1(t)$ from zero to its final value while remaining adiabatic when $M_z$ changed rapidly (but still fast enough so that metastable states survive, $\lesssim200 \Er/s$).  For points near the first-order phase transition three metastable states exist (Fig.~\ref{Fig:metastable}); near the second-order transition this count decreases, giving the two local minima which merge to a single minimum beyond the second-order transition.

We experimentally identified the number of metastable states by using $M_z$ and its higher moments, having started in each of the three $\ket{m_F = 0,\pm1}$ initial states.  A small variance indicates the final states are clustered together -- associated with a single global minimum in the free energy $G(M_z)$ -- and it increases when metastable or degenerate ground states are present. We distinguished systems with two degenerate magnetization states ($M_z\approx\pm1$) from those with three states by the same method, since when $M_z\approx\pm1$, the variance of $|M_z|$ is small, and it distinguishably increases as a third metastable state appears with $M_z=0$. In this way we fully mapped the system's metastable states in agreement with theory, as shown in Fig. \ref{Fig:metastable}.

%% file: Conclusion.tex

In conclusion, we accurately measured the two-parameter phase diagram of a spin-1 BEC, containing a ferromagnetic phase and an unmagnetized phase, continuously connecting a polar spinor BEC to a spin-helix BEC.  The ferromagnetic phase in this itinerant system is stabilized by SOC, and vanishes as the SOC strength $\hbar \kr$ goes to zero.   Our observation of controlled quench dynamics through a first-order phase transition opens the door for realizing Kibble-Zurek physics\cite{Kibble1976,Zurek1985} in this system, where the relevant parameters can be controlled at the individual $\rm Hz$ level.  The quadrupole tensor field $\propto \hat F_{zz}^{(2)}$ studied here is the $q=0$ component of the rank-2 spherical tensor operator $\hat F_q^{(2)}$, with $q\in\left\{\pm2,\pm1,0\right\}$.  The physics of this system would be further enriched by the addition of the remaining four tensor fields: this is straightforward using a combination of radio-frequency and microwave fields\cite{Nicklas2014}.  Furthermore, the first-order phase transition revealed in this work is further modified by the addition of spin-dependent interactions: very weak in $\Rb87$, giving effects below our current ability to resolve.  Our numerical studies show that a number of new symmetry-broken spinor phases emerge along the line that defines the first-order transition.  



%% file: MethodsSummary.tex

\begin{methodssummary}

\subsection{System Preparation}
We created $N\approx 4\times 10^5$ atoms $^{87}{\rm Rb}$ BECs in the ground electronic state $\ket{f=1}$ manifold\cite{Lin2009}, confined in the locally harmonic trapping potential formed at the intersection of two $1,064\ {\rm nm}$ laser beams propagating along ${\bf e}_x$ and ${\bf e}_y$ giving trap frequencies of $(\omega_x,\omega_y,\omega_z)/2\pi=(33(2),33(2),145(5))\ {\rm Hz}$.  The quadratic contribution to the $B_0=35.468(1)\ {\rm G}$ magnetic field's $\approx h\times25\ {\rm MHz}$ Zeeman shift lifts the degeneracy between the $\ket{f\shorteq1,m_F\shorteq-1}\leftrightarrow \ket{f\shorteq1,m_F\shorteq0}$ and $\ket{f\shorteq1,m_F\shorteq0}\leftrightarrow \ket{f\shorteq1,m_F\shorteq1}$ transitions, by $\epsilon = h\times90.417(1)\ {\rm kHz}$.   We denote the energy differences between these states as $\hbar\delta \omega_{-1,0}$, $\hbar\delta \omega_{0,+1}$, and $\hbar\delta \omega_{-1,+1}$.

\subsection{Frequency selective Raman coupling}
We Raman-coupled the three $m_F$ states using a pair of $\lambda=790.024(5)\ {\rm nm}$ laser beams counter-propagating along ${\bf e}_x$.  The beam traveling along $+{\bf e}_x$ had frequency components $\omega_{-1}^+$ and $\omega_{+1}^+$, while the beam traveling along $-{\bf e}_x$ contained the single frequency $\omega^-$.  These frequencies were chosen such that the differences $\delta\omega_{-1,0} \approx \omega^- - \omega_{-1}^+$ and $\delta\omega_{0,+1} \approx\omega^- - \omega_{+1}^+$ independently Raman coupled the $\ket{m_F=-1,0}$ and $\ket{m_F=0,+1}$ state-pairs, respectively.  Furthermore we selected $2\omega^- - (\omega_{-1}^+ +\omega_{+1}^+) = \delta \omega_{-1,+1}$ such that, after making the rotating wave approximation (RWA) the $\ket{m_F = \pm1}$ states were energetically shifted by the same $\Omega^{(0)}_2 = [(\delta\omega_{-1,0}-\delta\omega_{0,+1}) + (\omega_{-1}^+ - \omega_{+1}^+)]/2$ energy from $\ket{m_F = 0}$, thereby yielding the frequency-tuned tensor energy shift depicted in Fig.~\ref{Fig:Setup}{\bf a}. In addition, $\Omega_2$ in Eq.~\eqref{eq:Magnetic} differs from $\Omega^{(0)}_2$ by a small shift $\propto \Omega_1^2/\epsilon$ resulting from off-resonant coupling to transitions detuned by $2\epsilon$, which we computed directly using Floquet theory [see Eq.~(6) in Methods].  This tensor contribution $\Omega_2 \hat F_{zz}^{(2)}$ to the Hamiltonian might equivalently be introduced by a quadratic Zeeman shift alone, giving $\Omega_2\propto B_0^2>0$.  
\end{methodssummary}